\documentclass[journal]{IEEEtran}

\usepackage{amsmath}
\usepackage{bm}
\usepackage{hyperref}

\usepackage{graphicx}
\usepackage{glossaries}
\usepackage{hyperref}

\usepackage{booktabs}
\usepackage{makecell}

\usepackage{cleveref}
\Crefname{figure}{Fig.}{Figs.}
\Crefname{section}{Sec.}{Secs.}

\glsdisablehyper

\newacronym{aa}{AA}{Acoustic Acquisition}
\newacronym{ail}{AIL}{Alamire Interactive Lab}
\newacronym{allrad}{ALLRAD}{All-Round Ambisonic Decoding}
\newacronym{art}{ART}{Acoustic Ray Tracing}
\newacronym{asw}{ASW}{Apparent Source Width}
\newacronym{cad}{CAD}{Computer Aided Design}
\newacronym{cpu}{CPU}{Central Processing Unit}
\newacronym{cuda}{CUDA}{Compute Unified Device Architecture}
\newacronym{cvp}{CVP}{Consensus Vocabulary Profiling}
\newacronym{dirac}{DirAC}{Directional Audio Coding}
\newacronym{dmfa}{DMFA}{Dual Multiple Factor Analysis}
\newacronym{doa}{DOA}{Direction of Arrival}
\newacronym{dof}{DOF}{Degrees of Freedom}
\newacronym{drr}{DRR}{Direct-to-Reverberant Ratio}
\newacronym{edt}{EDT}{Early Decay Time}
\newacronym{ess}{ESS}{Exponential Sine Sweep}
\newacronym{fcp}{FCP}{Free Choice Profiling}
\newacronym{fcr}{FCR}{Feedback Canceling Reverberator}
\newacronym{fdl}{FDL}{Frequency-Domain Delay-Line}
\newacronym{fdn}{FDN}{Feedback Delay Network}
\newacronym{fft}{FFT}{Fast Fourier Transform}
\newacronym{fir}{FIR}{Finite Impulse Response}
\newacronym{foa}{FOA}{First-Order Ambisonics}
\newacronym{fp}{FP}{Flash Profile}
\newacronym{fwht}{FWHT}{Fast Walsh-Hadamard Transform}
\newacronym{gcc}{GCC}{Generalized Cross-Correlation}
\newacronym{gpa}{GPA}{Generalized Procrustes Analysis}
\newacronym{gpu}{GPU}{Graphics Processing Unit}
\newacronym{hoa}{HOA}{Higher-Order Ambisonic}
\newacronym{hrtf}{HRTF}{Head-Related Transfer Function}
\newacronym{hws}{HWS}{Historical Worship Spaces}
\newacronym{icc}{ICC}{Interaural Cross-Correlation}
\newacronym{ifft}{IFFT}{Inverse Fast Fourier Transform}
\newacronym{iir}{IIR}{Infinite Impulse Response}
\newacronym{imu}{IMU}{Inertial Measurement Unit}
\newacronym{ir}{IR}{Impulse Response}
\newacronym{ism}{ISM}{Image-Source Method}
\newacronym{ivp}{IVP}{Individual Vocabulary Profiling}
\newacronym{jit}{JIT}{Just-In-Time}
\newacronym{jnd}{JND}{Just Noticeable Difference}
\newacronym{lti}{LTI}{Linear Time-Invariant}
\newacronym{mdap}{MDAP}{Multiple-Direction Amplitude Panning}
\newacronym{mfa}{MFA}{Multiple Factor Analysis}
\newacronym{midi}{MIDI}{Musical Instrument Digital Interface}
\newacronym{mls}{MLS}{Maximum Length Sequence}
\newacronym{mushra}{MUSHRA}{MUltiple Stimuli with Hidden Reference and Anchor}
\newacronym{nls}{NLS}{Nearest Loudspeaker Synthesis}
\newacronym{nupc}{NUPC}{Non-Uniform Partitioned Convolution}
\newacronym{ola}{OLA}{Overlap-Add}
\newacronym{ols}{OLS}{Overlap-Save}
\newacronym{pca}{PCA}{Principal Component Analysis}
\newacronym{pe}{PE}{Perceptual Evaluation}
\newacronym{qda}{QDA}{Quantitative Descriptive Analysis}
\newacronym{rfft}{RFFT}{Real Fast Fourier Transform}
\newacronym{rir}{RIR}{Room Impulse Response}
\newacronym{rt}{RT}{Reverberation Time}
\newacronym{rta}{RTA}{Real-Time Auralization}
\newacronym{sdm}{SDM}{Spatial Decomposition Method}
\newacronym{sdn}{SDN}{Scattering Delay Network}
\newacronym{sfa}{SFA}{Sound Field Analysis}
\newacronym{sfs}{SFS}{Sound Field Synthesis}
\newacronym{shd}{SHD}{Spherical Harmonic Decomposition}
\newacronym{snr}{SNR}{Signal-to-Noise Ratio}
\newacronym{srir}{SRIR}{Spatial Room Impulse Response}
\newacronym{sti}{STI}{Speech Transmission Index}
\newacronym{svd}{SVD}{Singular Value Decomposition}
\newacronym{vbap}{VBAP}{Vector Base Amplitude Panning}
\newacronym{vr}{VR}{Virtual Reality}
\newacronym{wfa}{WFA}{Wave Field Analysis}
\newacronym{wfs}{WFS}{Wave Field Synthesis} 

\begin{document}

\title{Accelerated Interactive Auralization of Highly Reverberant Spaces using Graphics Hardware}

\author{\IEEEauthorblockN{Hannes Rosseel and Toon van Waterschoot} \\
    \IEEEauthorblockA{\textit{KU Leuven, Dept. of Electrical Engineering (ESAT), STADIUS Center for Dynamical Systems,\\ Signal Processing and Data Analytics, Leuven, Belgium}\\
        \{firstname.lastname\}@esat.kuleuven.be}
}

\maketitle
 
\begin{abstract}
Interactive acoustic auralization allows users to explore virtual acoustic environments in real-time, enabling the acoustic recreation of concert hall or \gls{hws} that are either no longer accessible, acoustically altered, or impractical to visit. Interactive acoustic synthesis requires real-time convolution of input signals with a set of synthesis filters that model the space-time acoustic response of the space. The acoustics in concert halls and \gls{hws} are both characterized by a long reverberation time, resulting in synthesis filters containing many filter taps. As a result, the convolution process can be computationally demanding, introducing significant latency that limits the real-time interactivity of the auralization system. In this paper, the implementation of a real-time multichannel loudspeaker-based auralization system is presented. This system is capable of synthesizing the acoustics of highly reverberant spaces in real-time using \gls{gpu}-acceleration. A comparison between traditional \gls{cpu}-based convolution and \gls{gpu}-accelerated convolution is presented, showing that the latter can achieve real-time performance with significantly lower latency. Additionally, the system integrates acoustic synthesis with acoustic feedback cancellation on the \gls{gpu}, creating a unified loudspeaker-based auralization framework that minimizes processing latency.
\end{abstract}

\glsresetall
\section{Introduction}
\label{sec:gpu2025:introduction}

\noindent Room acoustic auralization has been a topic of interest in audio engineering for several decades \cite{valimaki2012fifty, valimaki2016more,vorlander2020auralization}. Auralization refers to the process of recreating an acoustic environment through sound reproduction techniques. This is typically achieved by convolving an audio signal with a set of auralization filters that model the acoustic characteristics of a given space. When the output of this convolution process is played back through loudspeakers or headphones, the listener perceives the sound as if it was recorded in the modeled acoustic environment \cite{vorlander2020auralization}. Applications of room acoustic auralization span various domains, including architectural acoustics, virtual reality, immersive art installations, and music production, among others \cite{vorlander2020auralization}.

Interactive auralization extends these principles by enabling real-time exploration of the reproduced acoustic environment, allowing users to experience and interact with the various acoustic environments through live audio input. For vocal musicians and researchers of early music, interactive auralization provides a valuable tool to explore the intrinsic connection between the performance of music and the acoustics of the space for which the music was composed \cite{schiltz2003church, ueno2003experimental, ueno2010effect}. This is particularly relevant for concert halls and \gls{hws}, where interactive auralization facilitates the preservation and study of the unique acoustic properties of these culturally significant locations \cite{bassuet2004acoustics}.

An important consideration of interactive auralization systems is latency, as significant delays between user interactions and system responses can degrade the sense of immersion and responsiveness \cite{lindau2009perception, vorlander2020auralization}. To ensure real-time performance, auralization systems must perform the acoustic synthesis and any other processing in the sound acquisition-reproduction loop with minimal latency, ensuring that the synthesized sound field is perceived by the users without noticeable delay. The maximum allowable delay between the input audio signal and the resulting synthesized sound field is referred to as the latency budget \cite{vorlander2020auralization}. For block-based processing, the latency budget is defined as the buffer latency $n_x / f_s$, where $n_x$ denotes the block size in samples and $f_s$ is the operating sampling frequency of the auralization system. For sample-based processing, the buffer latency is $1 / f_s$, requiring the system to process each sample within a single sample period.

The modeling of rich and complex acoustics, such as those found in concert halls and \gls{hws}, require many filter taps to accurately capture the long reverberation times and spatial characteristics of the space. During real-time auralization, the convolution of input audio with these filters can be computationally expensive, especially in multichannel auralization systems containing many loudspeakers, as each loudspeaker is associated with its own synthesis filter. This makes meeting the desired latency budget, which is typically in the order of a few milliseconds, challenging to achieve using traditional consumer-grade hardware.

Moreover, in loudspeaker-based auralization systems that allow for interactive exploration of virtual acoustic environments, the presence of acoustic feedback can introduce additional challenges. Acoustic feedback can occur when the microphone capturing input audio is positioned near the loudspeakers used to reproduce the synthesized sound field. When the synthesized sound field is picked up by the microphone and reintroduced into the system, a closed feedback loop is formed. This can cause system instability, leading to undesirable artifacts such as howling and coloration of the synthesized sound \cite{vanwaterschoot2011fifty}. To mitigate this issue in real-time interactive auralization systems, where the feedback path between the loudspeakers and the microphone is relatively time-invariant, feedback cancellation algorithms that identify and cancel the feedback signal have been proposed \cite{abel2018feedback}. It should be noted that in these algorithms, it is assumed that the feedback path between the loudspeakers and the microphones is known a priori and remains relatively time-invariant during the operation of the system.

Traditional implementations of real-time auralization systems are often constrained by the computational complexity of the convolution process, and therefore limited in the number of channels that can be processed in real-time. To address this challenge, \gls{gpu} acceleration has been proposed as a means to reduce the processing latency for multichannel auralization by executing the convolution operations in parallel \cite{wefers2010high, savioja2010use}. Prior work has demonstrated the effectiveness of \gls{gpu} acceleration for real-time binaural auralization \cite{cowan2008spatial, cowan2009gpu, tsingos2009using, mauro2011binaural, belloch2012headphone, belloch2013headphone}, multichannel audio systems \cite{belloch2011real}, and \gls{wfs} auralization \cite{belloch2013gpu, ranjan2014fast, belloch2017gpu}. Notably, Belloch et al. \cite{belloch2017gpu} implemented a real-time \gls{gpu}-based \gls{wfs} system with room compensation that is capable of auralizing complex sound fields with moving sources. This work demonstrated the feasibility of using \gls{gpu} acceleration for real-time auralization and highlighted the potential benefits of \gls{gpu} acceleration for reducing latency and improving computational efficiency in auralization systems. However, the integration of feedback cancellation algorithms with \gls{gpu}-based convolution for real-time interactive auralization has not been extensively explored in the literature.

This paper aims to bridge this gap by presenting a generalizable framework for real-time multichannel auralization systems that integrates uniform partitioned convolution \cite{wefers2014partitioned, wefers2010high} with real-time feedback cancellation, and utilizes \gls{gpu} acceleration to minimize processing latency. Compared to previous \gls{gpu}-based approaches limited to specific auralization methods, such as \gls{wfs} \cite{belloch2013gpu, ranjan2014fast, belloch2017gpu}, the proposed framework supports scalable loudspeaker-based synthesis with practical flexibility in filter length and channel count. This enables the acoustic synthesis of complex acoustic environments that exhibit long reverberation times, including concert hall and historical worship space acoustics, while maintaining low-latency performance.

To validate the effectiveness of the proposed framework, a performance evaluation is conducted which compares the latency performance of the \gls{gpu}-accelerated implementation to a \gls{cpu}-based implementation. The results demonstrate that the proposed framework can achieve low-latency real-time auralization of \gls{hws} acoustics with minimal system instability for large filter lengths and channel counts. This work highlights the potential of \gls{gpu} acceleration for reducing latency and improving computational efficiency in real-time interactive auralization systems, while maintaining system stability in the presence of acoustic feedback.

The main contributions of this paper can be summarized as follows:\vspace{-1em}
\begin{itemize}
    \item A generalizable framework for real-time multichannel auralization systems that integrates uniform partitioned convolution with real-time feedback cancellation, utilizing \gls{gpu} acceleration to minimize processing latency.
    \item A performance evaluation comparing the \gls{gpu}-based implementation with a \gls{cpu}-based implementation in terms of processing latency for various filter lengths, channel counts, and block sizes. Additionally, the end-to-end latency is reported for different block sizes.
    \item The public release of the source code for the proposed framework to support further research and development in interactive auralization.
\end{itemize}

The paper is organized as follows. In \Cref{sec:gpu2025:notation}, the mathematical notation used throughout the paper is introduced. \Cref{sec:gpu2025:interactive_multichannel_auralization} provides an overview of the proposed system implementation, including a detailed description of fast convolution methods. In \Cref{sec:gpu2025:uniform_partitioned_convolution}, the uniform partitioned convolution algorithm used in the proposed framework for real-time auralization is presented. \Cref{sec:gpu2025:implementation} details the Python implementation of the partitioned convolution algorithm and the real-time auralization system. In \Cref{sec:gpu2025:performance}, the performance benchmark results of the proposed framework is presented, which compares the processing time of the \gls{gpu}-based implementation with the \gls{cpu}-based implementation. In \Cref{sec:gpu2025:e2e_latency}, the end-to-end latency of the proposed auralization system is presented for different block sizes. Finally, \Cref{sec:gpu2025:conclusion} concludes the paper and discusses future research directions.

\section{Mathematical notation}
\label{sec:gpu2025:notation}

\noindent In this paper, scalar quantities are denoted by italic letters (e.g., $a$ or $A$), while vectors are represented by bold lowercase letters (e.g., $\mathbf{a}$). Matrices are denoted by bold uppercase letters (e.g., $\mathbf{A}$). Quantities that vary over time are denoted by an argument in parentheses (e.g., $a(t)$ or $\mathbf{a}(t)$).

\section{Interactive multichannel auralization}
\label{sec:gpu2025:interactive_multichannel_auralization}

\noindent Interactive multichannel auralization systems aim to reproduce the acoustics of a virtual or modelled environment around a listener in real-time. In this paper, the aim is to develop an interactive multichannel auralization system that can be used as a rehearsal space for vocal performers and musicians. A system with similar functionality, implemented using commercial hardware, was used in \cite{rosseel2023interactive}. Interactive refers to the ability of a system to react to user inputs in real-time. In the context of auralization systems, this means that the system processes live audio input from microphones, allowing the users to receive immediate auditory feedback that simulates the acoustics of a specific environment. This real-time interaction enables musicians to adapt their performance based on the acoustic characteristics of the simulated space \cite{amengual2017investigations, amengual2019analysis,rosseel2023interactive}.

Since the setup of the system is intended to be operated by vocal performers and musicians, which are not necessarily experts in acoustics or signal processing, the system should be straightforward to set up without the need for installing additional equipment. Therefore, in this paper, the focus lies on static loudspeaker-based interactive auralization systems, as opposed to headphone-based binaural auralization systems.

\subsection{System overview}

\begin{figure}[t]
    \centering
    \includegraphics[width=\columnwidth]{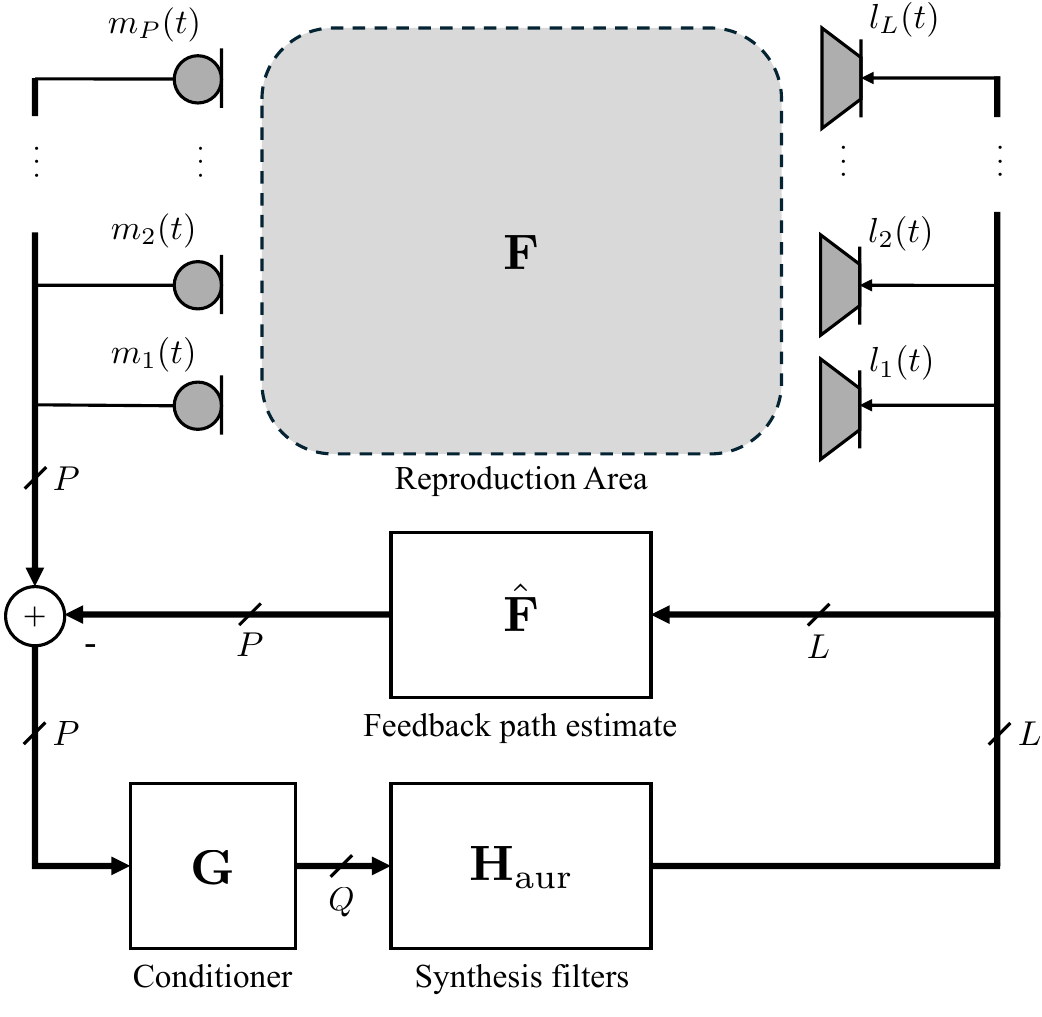}
    \caption{Block diagram of an interactive multichannel auralization system containing $L$ loudspeakers and $P$ microphones. The system consists of a pre-processing stage, where the microphone signals are processed by a pre-processor function $\mathbf{G}$, a synthesis stage with synthesis filters $\mathbf{H}_\textrm{aur}$, and an acoustic feedback cancellation stage which subtracts the estimated feedback signals from the microphone input. The feedback path for each loudspeaker-microphone pair is estimated a-priori and stored in the estimated feedback path matrix $\mathbf{\hat{F}}$.}
    \label{fig:gpu2025:interactive_auralization}
\end{figure}

The block diagram of an interactive loudspeaker-based multichannel auralization system is shown in \Cref{fig:gpu2025:interactive_auralization}. The system consists of a reproduction area that is surrounded by $L$ loudspeakers and $P$ microphones, which are used to reproduce the acoustics of the virtual environment and capture the input signals, respectively. The input signals $\mathbf{m}(t)$ are first conditioned using a pre-processing stage $\mathbf{G}$, a $P \times Q$ matrix of operators, which can include equalization, compression, or weighting for different input channels. The conditioned input signals in the $Q \times 1$ vector $\mathbf{\tilde{m}}(t)$ are then filtered with a set of $L$ synthesis filters, after which the resulting  $L \times 1$ output signals $\mathbf{l}(t)$ are played back by the loudspeakers. The output signals are described by

\begin{equation}
    \mathbf{l}(t) = \mathbf{H}_\textrm{aur}^T * \mathbf{\tilde{m}}(t) = \begin{bmatrix}
        \mathbf{h}_{11} & \mathbf{h}_{12} & \cdots & \mathbf{h}_{1L} \\
        \mathbf{h}_{21} & \mathbf{h}_{22} & \cdots & \mathbf{h}_{2L} \\
        \vdots & \vdots & \ddots & \vdots \\
        \mathbf{h}_{Q1} & \mathbf{h}_{Q2} & \cdots & \mathbf{h}_{QL}
    \end{bmatrix}^T * \begin{bmatrix}
        \tilde{m}_1(t) \\
        \tilde{m}_2(t) \\
        \vdots \\
        \tilde{m}_Q(t)
    \end{bmatrix},
\end{equation}

\noindent where $\mathbf{H}_\textrm{aur}$ is a $Q \times L$ matrix of linear filter operators, $(\cdot)^T$ is the transpose operator, and $(\ast)$ denotes the convolution operator. The output signal for loudspeaker $l$ is given by:

\begin{equation}
    l_l(t) = \sum_{q=1}^{Q} \mathbf{h}_{ql} * \tilde{m}_q(t).
\end{equation}

The synthesis filters $\mathbf{H}_\textrm{aur}$ are designed to render the acoustics of the virtual or modeled environment inside the reproduction area. These filters are the result of applying \gls{sfa} and \gls{sfs} methods to a measured or modeled \gls{rir} \cite{vorlander2020auralization, rosseel2025state}. For an overview of \gls{sfa} and \gls{sfs} methods for auralization, the reader is referred to \cite{rosseel2025state}.

In loudspeaker-based auralization systems, each loudspeaker and microphone pair has an acoustic feedback path $\mathbf{f}_{lp}$, which models the acoustic path from loudspeaker $l$ to microphone $p$. The collection of all acoustic feedback paths in the auralization system can be represented by the acoustic feedback matrix $\mathbf{F}$, which is a $P \times L$ matrix of linear filter operators. To prevent acoustic feedback in the system, acoustic feedback cancellation can be achieved by first estimating the acoustic feedback path matrix $\mathbf{\hat{F}}$, which models the acoustic path between all microphone and loudspeaker pairs. Subsequently, the $P \times 1$ feedback signals $\mathbf{c}(t)$ can be obtained by filtering the output signals $\mathbf{l}(t)$ with the estimated acoustic feedback path matrix $\mathbf{\hat{F}}$, as

\begin{equation}
    \mathbf{c}(t) = \mathbf{\hat{F}} * \mathbf{l}(t) = \begin{bmatrix}
        \mathbf{\hat{f}}_{11} & \mathbf{\hat{f}}_{12} & \cdots & \mathbf{\hat{f}}_{1L} \\
        \mathbf{\hat{f}}_{21} & \mathbf{\hat{f}}_{22} & \cdots & \mathbf{\hat{f}}_{2L} \\
        \vdots & \vdots & \ddots & \vdots \\
        \mathbf{\hat{f}}_{P1} & \mathbf{\hat{f}}_{P2} & \cdots & \mathbf{\hat{f}}_{PL}
    \end{bmatrix} * \begin{bmatrix}
        l_1(t) \\
        l_2(t) \\
        \vdots \\
        l_L(t)
    \end{bmatrix}.
\end{equation}

The acoustic feedback is then canceled by subtracting the feedback signals from the microphone signals. Various methods for estimating the acoustic feedback path matrix $\mathbf{\hat{F}}$ have been proposed in the literature \cite{vanwaterschoot2011fifty, abel2018feedback}.

To allow interactive auralization of virtual or modeled acoustics in real-time, the system should be able to perform two main operations under real-time constraints: (1) the convolution of the input signals with the synthesis filters $\mathbf{H}_\textrm{aur}$, and (2) the convolution of the output signals $\mathbf{l}(t)$ with the estimated acoustic feedback path matrix $\mathbf{\hat{F}}$. The processing latency of the system should be kept as low as possible to ensure a seamless user experience.

\subsection{Real-time convolution}

The convolution of the input signals with the synthesis filters $\mathbf{H}_\textrm{aur}$ and the convolution of the output signals $\mathbf{l}(t)$ with the estimated acoustic feedback path matrix $\mathbf{\hat{F}}$ are computationally intensive operations that need to be performed within a predetermined latency budget. For large synthesis filters and many output channels, the convolution process can become infeasible for real-time applications on consumer-grade hardware. To reduce the computational complexity of the convolution process, various methods have been proposed in the literature.

Multichannel block-based time-domain convolution requires a number of operations that is proportional to $\mathcal{O}(n \cdot n_x \cdot n_h)$, where $n$, $n_x$ and $n_h$ are the number of channels, the block size and the length of the filter, respectively. The input block is defined as $\mathbf{x} = [x(t - n_x T), \ldots, x(t - T)]$, where $T = 1 / f_s$ is the sampling period and $f_s$ is the sampling frequency. The time-domain convolution for a single-channel input block $\mathbf{x}$ of length $n_x$ and a single-channel filter $\mathbf{h}$ of length $n_h$ is given by:
\begin{equation}
y[\kappa] = \sum_{m=1}^{n_h} h[m] \cdot x[\kappa - m], \quad \kappa = 1, 2, \ldots, n_x + n_h - 1,
\end{equation}

\noindent where $\mathbf{y}$ is the resulting output block of length $n_x + n_h - 1$, and the appropriate zero-padding is applied to the input block $\mathbf{x}$ for $\kappa - m < 0$.

To accelerate the convolution process, various methods have been proposed in the literature, such as the use of block-based frequency-domain convolution using the \gls{fft} \cite{cooley1965algorithm}. Block-based frequency-domain convolution can be used to reduce the number of operations required for convolution to $\mathcal{O}(n \cdot n_f \log n_f)$, where $n_f \ge n_h + n_x - 1$ is the \gls{fft}-size. First, the input block $\mathbf{x}$ and the filter $\mathbf{h}$ are zero-padded and transformed into the frequency-domain using an $n_f$-point \gls{fft}. The resulting frequency spectra $\mathbf{\bar{x}}$ and $\mathbf{\bar{h}}$ are then multiplied element-wise, after which the resulting spectrum is transformed back into the time-domain using an $n_f$-point \gls{ifft} to obtain the output $\mathbf{y}$. The block-based frequency-domain convolution for a single-channel input block $\mathbf{x}$ and a single-channel filter $\mathbf{h}$ is given by:
\begin{equation}
    \mathbf{y} = \text{IFFT}(\mathbf{\bar{h}} \odot \mathbf{\bar{x}}),
\end{equation}

\noindent where $(\odot)$ denotes the element-wise multiplication operator between two vectors.

While block-based frequency-domain convolution can significantly reduce the computational complexity of the convolution process as compared to time-domain convolution, it introduces additional processing due to the \gls{fft}-size being lower-bounded by $n_h + n_x - 1$ frequency bins. For long filter sizes $n_h$, this additional processing can introduce latency into the system that could exceed the buffer latency $n_x / f_s$, which, for real-time auralization systems, is usually only a few milliseconds long.

To further reduce the latency of block-based frequency-domain convolution with long filter lengths $n_h$, partitioned convolution methods have been proposed in the literature \cite{stockham1966high, torger2001realtime, wefers2014partitioned}. These methods decompose the filter into smaller sub-filters that can be processed independently using block-based frequency-domain convolution. Partitioned convolution can be implemented in two ways: uniform partitioned convolution and non-uniform partitioned convolution.

In uniform partitioned convolution, the filter of length $n_h$ is divided into $K$ partitions or sub-filters of length $n_k$, such that $n_h \ge K \cdot n_k$. An input block $\mathbf{x}$ with block-length $n_x$ is then convolved with each sub-filter in the frequency-domain using the \gls{fft}. Each sub-filter $\mathbf{h}_k$ is offset by $k\cdot n_k$ samples, where $k$ denotes the index of the sub-filter. Moreover, when $n_x = n_k$, this delay can be directly realized in the frequency-domain \cite{wefers2014partitioned}. The resulting single-channel output $\mathbf{y}$ from convolving an input block $\mathbf{x}$ with a filter $\mathbf{h}$ using uniform partitioned convolution is given by:
\begin{equation}
    \mathbf{y} = \sum_{k=1}^{K} \text{IFFT}\left(\mathbf{\bar{h}}_k \odot \mathbf{\bar{x}}\right) * \delta[\kappa - (k-1) \cdot n_k],
\end{equation}

\noindent where $\mathbf{\bar{h}}_k$ denotes the frequency response of sub-filter $\mathbf{h}_k$ and $\delta[\cdot]$ is the Kronecker delta function which realizes the sample delay of each sub-filter in the time-domain.

Non-uniform partitioned convolution is similar to uniform partitioned convolution, but the filter is divided into $K$ sub-filters of different lengths. This allows for a more efficient use of the available latency budget, as an optimal length can be chosen for each sub-filter to fit within the available latency budget. However, non-uniform partitioned convolution introduces additional complexity due to the varying sub-filter lengths, causing the delay of each sub-filter to be realized in the time domain \cite{wefers2014partitioned}. In this paper, only the uniform partitioned convolution method is considered, as it is easier to implement and does not require additional complexity in the implementation.

Partitioned convolution methods are well-suited for parallel processing on multicore processors since the sub-filters can be processed independently of each other. This makes partitioned convolution methods particularly suitable for implementation on \gls{gpu} architectures. These architectures are designed to perform parallel processing of large amounts of data, at magnitudes higher floating-point performance compared to traditional multicore \gls{cpu} architectures. In \cite{wefers2010high}, a uniform partitioned convolution method was proposed for real-time auralization systems implemented on a \gls{gpu} architecture. The implementation was shown to be capable of processing large numbers of sub-filters, while maintaining low latency and high computational efficiency. 

Building upon the work of \cite{wefers2010high}, this paper proposes a novel extension by combining acoustic synthesis and feedback cancellation into a unified implementation that employs uniform partitioned convolution on the \gls{gpu}. This approach advances the original implementation, which only focused on the synthesis stage. By integrating feedback cancellation into the same framework, a more efficient and streamlined processing pipeline can be achieved. In the following section, a detailed description of the uniform partitioned convolution algorithm \cite{wefers2010high} is provided, which serves as the foundation for the proposed implementation.

\section{Uniform partitioned convolution}
\label{sec:gpu2025:uniform_partitioned_convolution}

\begin{figure*}[t]
    \centering
    \includegraphics[width=\textwidth]{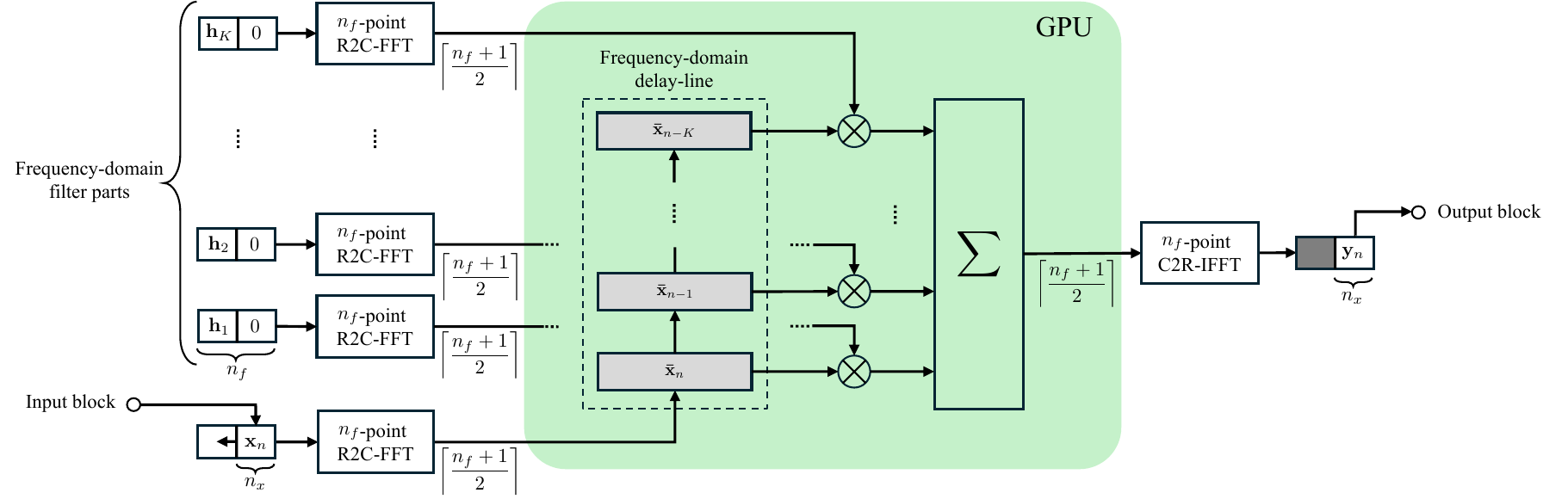}
    \caption{Block diagram of the uniform partitioned convolution algorithm used in \cite{wefers2010high}. The algorithm is visualized for a single-channel input block $\mathbf{x}_n$ and output block $\mathbf{y}_n$. The \gls{fft}-size for the input and filters was set to $n_f = 2 n_x$, where $n_x$ is the block size. The frequency-domain input blocks are denoted by $\mathbf{\bar{x}}_n$. The resulting output block is obtained using the \gls{ols} method. The multiplication and summation of the input block spectrum with each of the sub-filter responses can be performed entirely in the frequency-domain on the \gls{gpu}.}
    \label{fig:gpu2025:uniform_partitioned_convolution}
\end{figure*}

\noindent The block diagram of the uniform partitioned convolution algorithm used in \cite{wefers2010high} is shown in \Cref{fig:gpu2025:uniform_partitioned_convolution}. The algorithm is visualized for a single-channel input block $\mathbf{x}_n$ and output block $\mathbf{y}_n$, but can be trivially extended to multiple channels. During initialization, each filter in the matrix $\mathbf{H}_\textrm{aur}$ is split into $K$ sub-filters of length $n_k$. Typically, the length of each sub-filter is chosen such that $n_k = n_x$, where $n_x$ is the block size of the input signal, as this allows for the delay of each sub-filter to be directly realized in the frequency-domain \cite{wefers2010high}. As a result, the \gls{fft}-size for the input and filters is set to $n_f = 2 n_x$, which is the minimum \gls{fft}-size when $n_k = n_x$. Each sub-filter is then padded with an additional $n_x$ zeros, before being transformed into the frequency-domain using an $n_f$-point real-to-complex \gls{fft} to obtain $\mathbf{\bar{h}}_k$. The algorithm then comprises three main stages:

\begin{enumerate}
    \item \textbf{Input transformation:} For stream processing, each incoming input block $\mathbf{x}_n$ is copied at the end of an $n_f$-point sliding window buffer after left-shifting by $n_x$ samples. This is called input packing. This buffer is then transformed into the frequency-domain using the real-to-complex \gls{fft}. The resulting frequency spectrum is stored in a buffer of length $\lceil (n_f + 1) / 2\rceil$ frequency bins. Since this step does not benefit from parallel processing, it is always performed on the \gls{cpu}.
    \item \textbf{Frequency-domain convolution with sub-filters:} The frequency spectrum of the input block $\mathbf{\bar{x}}_n$ is stored into a \gls{fdl} buffer containing $K$ blocks. With each new input block, the contents in the \gls{fdl} are shifted up by one block. Hereafter, the frequency response of each sub-filter is multiplied with the corresponding frequency-domain block in the \gls{fdl} buffer. Since these operations are independent of each other, they can benefit from parallel processing on the \gls{gpu}. The resulting spectra are accumulated in an output frequency-domain buffer.
    \item \textbf{Output transformation:} The accumulated spectrum is transformed back into the time-domain using the complex-to-real \gls{ifft}, resulting in $n_f$ output samples. Due to \gls{ols}, only the last $n_x$ samples are valid as the output block $\mathbf{y}_n$. This is called output unpacking, and this step is performed on the \gls{cpu} as it does not benefit from parallel processing.
\end{enumerate}

By implementing the frequency-domain convolution for each sub-filter on a \gls{gpu} architecture, fast uniform partitioned convolution can be achieved when processing the input signals with the synthesis filters $\mathbf{H}_\textrm{aur}$, and when processing the output signals $\mathbf{l}(t)$ with the estimated acoustic feedback path matrix $\mathbf{\hat{F}}$. This unified approach leverages the parallel processing capabilities of modern \glspl{gpu} to significantly reduce the computational complexity of the convolution process, especially for long synthesis filters and large numbers of output channels. Moreover, it allows for real-time auralization of complex acoustic environments, such as those found in concert halls and \gls{hws}, while maintaining low latency and high computational efficiency.

\section{Python implementation}
\label{sec:gpu2025:implementation}

\noindent The implementation of the real-time auralization system is done in Python 3.13.1. While Python is generally not regarded as the most efficient language for real-time audio processing, it does allow for rapid prototyping and easy integration with various libraries that can be used to accelerate processing. Such accelerated libraries include \texttt{numpy}, \texttt{numba}, and \texttt{torch}, which offer wrapper functions around optimized C and \gls{cuda} code to enable real-time performance in Python. The code of this implementation is made available on GitHub\footnote{\url{https://github.com/hrosseel/GPU-accelerated-auralization}}, a popular open-source code hosting and collaboration platform.

The Python implementation consists of two main components: the uniform partitioned convolution algorithm outlined in \Cref{sec:gpu2025:uniform_partitioned_convolution} and the real-time auralization system that implements both acoustic synthesis and feedback cancellation. The auralization system uses the uniform partitioned convolution implementation to convolve the input signal with a set of time-domain synthesis filters, and the output signal with a set of time-domain feedback cancellation filters. The implementation is designed to be modular and extensible, allowing for easy integration with other components and systems. The code is structured in a way that allows for easy modification and extension, making it suitable for research and development purposes.

Uniform partitioned convolution is implemented in the \texttt{PartitionedConvolution} Python class. This class is initialized by a set of $C_\textrm{out}$ time-domain filters, a block size $n_x$, a number of input channels $C_\textrm{in}$, an \gls{fft} size $n_f$, and a \texttt{device} parameter that specifies whether the processing should be done on the \gls{cpu} or \gls{gpu}. The class provides a \texttt{convolve} method that inputs a block of samples of size $(C_\textrm{in}, n_x)$, and returns a block of filtered output samples of size $(C_\textrm{out}, n_x)$. Parameter $C_\textrm{in}$ can be set to either $1$ or $C_\textrm{out}$. If $C_\textrm{in} = 1$, the single input channel is filtered independently with each of the $C_\textrm{out}$ filters, resulting in $C_\textrm{out}$ output channels. If $C_\textrm{in} = C_\textrm{out}$, the input signal is filtered with the corresponding filter, resulting in a one-to-one filtering of the input channels with the filters.

The real-time auralization system is implemented in the \texttt{PartitionedAuralization} Python class. This class is initialized by a set of $C_\textrm{out}$ time-domain synthesis filters, a set of corresponding time-domain feedback cancellation filters, a block size $n_x$, an \gls{fft} size $n_f$, and a \texttt{device} parameter that specifies whether the processing should be done on the \gls{cpu} or \gls{gpu}. The class provides an \texttt{auralize} method that takes a block of input samples of size $(1, n_x)$ as input, and returns a block of auralized output samples of size $(C_\textrm{out}, n_x)$ that can be played back by a loudspeaker system in the reproduction area.

The \texttt{PartitionedAuralization} Python class's \texttt{auralize} method that internally uses the \texttt{convolve} method of the \texttt{PartitionedConvolution} class to perform the convolution of the input samples with the synthesis filters, and of the output samples with the feedback cancellation filters. This approach allows for efficient processing as the output of the filtering with the synthesis filters can be directly used as input to the filtering with the feedback cancellation filters, all without the need to move the data from the \gls{gpu} to the \gls{cpu}. The estimated feedback path signals are subtracted from the next block of input samples to cancel the contribution of the loudspeaker signals from the input signal.

The current implementation uses a default \gls{fft} size of $n_f = 2 n_x$. It should be noted that the output of the input block $\mathbf{x}_n$ filtered with the synthesis filters $\mathbf{H}_\textrm{aur}$ could be used directly as frequency-domain input to the filtering operation with the feedback cancellation filters. However, while a computational saving of $C_\textrm{out}$ \gls{fft} operations could be achieved, this approach does require a change in \gls{fft} size from $n_f = 2 n_x$ to $n_f = 3 n_x$. This increase in \gls{fft}-size would result in a computational increase of $(K_\textrm{aur} + K_\textrm{fc}) \, C_\textrm{out} \, n_x$ complex multiplications, where $K_\textrm{aur}$ and $K_\textrm{fc}$ are the number of filter parts for the synthesis and feedback cancellation filters, respectively. This additional cost outweighs the potential saving of $C_\textrm{out}$ $n_f$-point \gls{fft} operations when $K_\textrm{aur} + K_\textrm{fc} \ge n_x \log(2 n_x)$, which is the case for large synthesis filters.

The performance gain of the \gls{gpu} implementation is evaluated by comparing the performance to a \gls{cpu} implementation of the algorithm. The \gls{cpu} implementation uses the \texttt{numba} library to accelerate the convolution operation. The \texttt{numba} library is a \gls{jit} compiler that translates Python code to optimized machine code at runtime. The \gls{cpu} implementation is run on multiple \gls{cpu} threads to further accelerate the processing. Similarly, the \gls{gpu} implementation of the \texttt{PartitionedConvolution} class uses the Python \texttt{torch} library to compile and call \gls{cuda} code. The \gls{cuda} code was written in a separate file and compiled using the \texttt{torch} library. A downside of using the \gls{cuda} framework instead of alternative frameworks, is that the portability of the implementation is limited to systems with \gls{cuda}-compatible \glspl{gpu}.

A limitation of the current implementation is that only a single input channel is supported for the auralization system. However, this limitation can be overcome by creating multiple instances of the \texttt{PartitionedAuralization} class in parallel and summing the respective output channels, thereby enabling support for multiple input channels.

\section{Performance benchmarking}
\label{sec:gpu2025:performance}

\noindent In this section, the performance of the implemented partitioned convolution algorithm and real-time auralization system is discussed. A performance comparison of the \gls{cpu} and \gls{gpu} implementations is presented as a function of block size, number of channels, and filter length, focusing on their impact on processing time.

The workstation used for the performance evaluation was equipped with an AMD Ryzen 9 5950X processor containing 16 \gls{cpu} cores and 32 threads, 16 GB of RAM, and an NVIDIA RTX A6000 \gls{gpu} containing 10.752 \gls{cuda} cores and 48 GB of \gls{gpu} memory. The \gls{gpu} is compatible with the \gls{cuda} 11.6 toolkit. The operating system of the workstation was Debian 12 with Linux kernel 5.10.0-26-amd64. In order to stabilize the performance measurements, the system was configured to isolate 12 out of the 16 \gls{cpu} cores for the real-time auralization system. The \gls{cpu} isolation was done using the \texttt{cgroups} feature of the Linux kernel. The \gls{cpu} cores were isolated to prevent other processes from interfering with the performance measurements, resulting in a system configuration where the real-time auralization system had exclusive access to 12 \gls{cpu} cores and the \gls{gpu}. This led to stable performance measurements.

\begin{figure}[ht]
    \centering
    \includegraphics[width=\columnwidth]{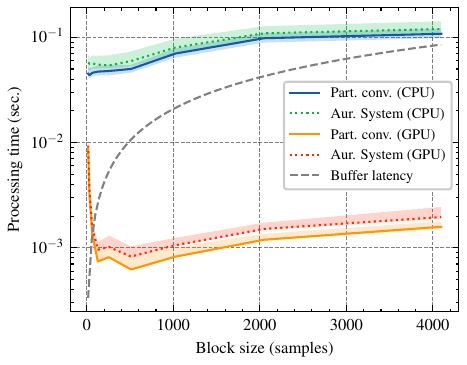}
    \caption{Performance of the \gls{cpu} and \gls{gpu} implementations of the partitioned convolution algorithm and the real-time auralization system as a function of the block size.}
    \label{fig:gpu2025:benchmark_block_size}
\end{figure}

\begin{figure}[ht]
    \centering
    \includegraphics[width=\columnwidth]{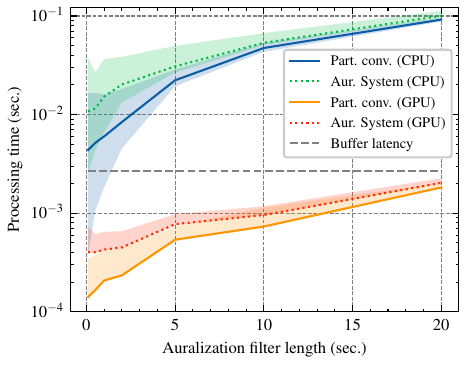}
    \caption{Performance of the \gls{cpu} and \gls{gpu} implementations of the partitioned convolution algorithm and the real-time auralization system as a function of the filter length.}
    \label{fig:gpu2025:benchmark_filter_length}
\end{figure}

The processing time of the \gls{cpu} and \gls{gpu} implementations of the partitioned convolution algorithm and the real-time auralization system were evaluated as a function of the block size $n_x$, the number of output channels $C_\textrm{out}$, and the filter length. The operating sampling rate of the auralization system was set to $48$ kHz, and the default block size was set to $n_x = 128$ samples, which corresponds to a buffer latency of $2.6$ ms. The default number of output channels was set to $C_\textrm{out} = 32$ channels, and the default filter length was set to 10 seconds, which corresponds with the length of auralization filters that model the acoustics of a highly reverberant space, e.g., a church. The performance evaluation was conducted by randomly generating the synthesis filters, feedback cancellation filters, and input samples from a standard normal distribution. The stochastic nature of the generated filters and input samples avoids potential performance biases that could arise from the use of specific filter coefficients or input samples.

System performance was evaluated by measuring the processing time required to filter a single block of input samples with a set of auralization filters, and optionally feedback cancellation filters. The filtering process consisted of three stages: transforming the input block to the frequency domain, performing frequency-domain uniform partitioned convolution, and converting the output block back to the time domain. In the \gls{gpu} implementation, this process additionally included the transfer of the frequency-domain input block to \gls{gpu} memory and the return of the processed frequency-domain output block to \gls{cpu} memory. All performance measurements were conducted after system initialization, with the filters pre-transformed to the frequency domain and loaded into either \gls{cpu} or \gls{gpu} memory.

The processing time was measured using the \texttt{time.perf\_counter()} function of the Python \texttt{time} module, and averaged over $10.000$ random trials to obtain a stable performance measurement. In the figures below, the processing time is shown as a function of the block size, auralization filter length, number of output channels, and feedback cancellation filter length. The mean processing time is shown as a solid line for the uniform partitioned convolution algorithm, and as a dashed line for the real-time auralization system implementation. The shaded area around the solid and dashed lines represents the maximum and minimum processing time measured over the $10.000$ random trials.

In \Cref{fig:gpu2025:benchmark_block_size}, the performance of the \gls{cpu} and \gls{gpu} implementations of the partitioned convolution algorithm and the real-time auralization system is shown as a function of the block size, which varied from $16$ to $4096$ samples corresponding to a buffer latency of $0.3$ ms to $85.3$ ms, respectively. To achieve real-time performance, the processing time of the real-time auralization system should be less than the buffer latency. It can be seen that for uniform partitioned convolution and auralization on the \gls{cpu}, real-time performance is not achieved for any of the block sizes. For the \gls{gpu} implementation of uniform partitioned convolution and auralization, real-time performance is achieved for block sizes higher than 128 samples. For these block sizes, the mean performance of the \gls{gpu} implementation is up to $60$ higher than the \gls{cpu} implementation.

In \Cref{fig:gpu2025:benchmark_filter_length}, the performance is shown as a function of the filter length, which is varied from $0.1$ to $20$ seconds. It can be seen that the processing time of both the \gls{cpu} and \gls{gpu} implementations increases as the filter length increases. With the \gls{cpu} implementation of the real-time auralization system, real-time filtering is not achieved, as the processing time exceeds the buffer latency for all filter lengths. However, the performance of the \gls{gpu} implementation is able to achieve real-time performance for filter lengths up to $20$ seconds.

\begin{figure}[ht]
    \centering
    \includegraphics[width=\columnwidth]{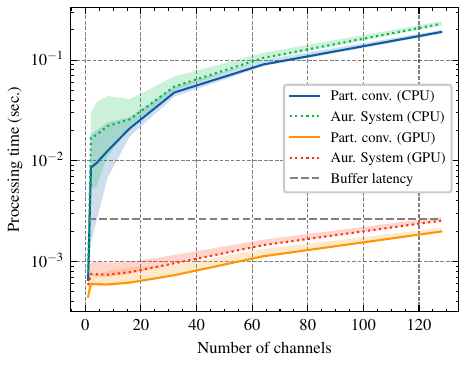}
    \caption{Performance of the \gls{cpu} and \gls{gpu} implementations of the partitioned convolution algorithm and the real-time auralization system as a function of the number of output channels.}
    \label{fig:gpu2025:benchmark_channels}
\end{figure}

\begin{figure}[ht]
    \centering
    \includegraphics[width=\columnwidth]{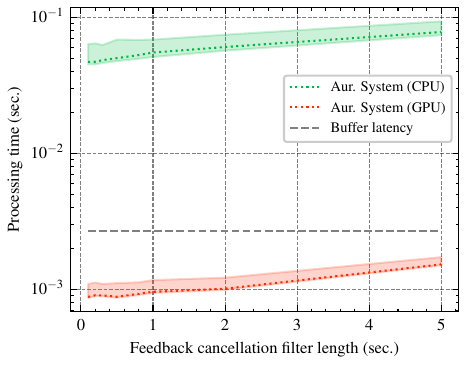}
    \caption{Performance of the \gls{cpu} and \gls{gpu} implementations of the partitioned convolution algorithm and the real-time auralization system as a function of the feedback cancellation filter length.}
    \label{fig:gpu2025:benchmark_filter_length_fc}
\end{figure}

The performance as a function of the number of output channels $C_\textrm{out}$ is shown in \Cref{fig:gpu2025:benchmark_channels}. The performance is evaluated for a number of output channels that is varied from $1$ to $128$ channels. It can be seen that the processing time of both the \gls{cpu} and \gls{gpu} implementations increases as the number of channels increases. However, the performance of the \gls{gpu} implementation decreases at a slower rate than the \gls{cpu} implementation. Real-time filtering is achieved for up to a single output channel on the \gls{cpu} and up to 120 output channels on the \gls{gpu}.

Finally, the performance is shown as a function of the feedback cancellation filter length in \Cref{fig:gpu2025:benchmark_filter_length_fc}, which is varied from $0.1$ to $5$ seconds. It can be seen that the overall performance of the real-time auralization system is not significantly impacted by the feedback cancellation filter length of up to $5$ seconds. This is due to the fact that the feedback cancellation filters are typically shorter than the synthesis filters.

In addition to the performance evaluation, the numerical precision of both implementations was assessed. Architectural differences in floating-point arithmetic and accumulation order can introduce small differences between the \gls{gpu} and \gls{cpu} implementations, which may accumulate over time. This is relevant for feedback cancellation, where precision drift can reduce cancellation effectiveness over time. To quantify the precision differences, validation tests were conducted using representative signals and filter configurations. The results showed that the \gls{gpu} closely matched the \gls{cpu} output, with maximum absolute error magnitudes on the order of $10^{-5}$.

\section{End-to-end latency}
\label{sec:gpu2025:e2e_latency}

\noindent In addition to the performance benchmarks discussed in \Cref{sec:gpu2025:performance}, the end-to-end latency of the proposed real-time auralization system was evaluated across a range of block sizes. End-to-end latency refers to the total delay between the incoming analog audio signal and the corresponding output signal. This includes analog-to-digital conversion, DSP processing chain, and digital-to-analog conversion. This latency is dependent on the specific hardware and software configuration, including the audio interface, operating system, and driver implementation. Therefore, the latencies presented in this section are specific to the tested configuration and may vary with different setups.

Measurements were conducted using an \textit{RME Fireface UFX+} audio interface connected via USB 3.0 to the Linux-based workstation described in \Cref{sec:gpu2025:performance}. The audio interface operated in Class-Compliant mode, and its input and output ports were physically looped back to form a closed signal path. The end-to-end latency values were measured at a sampling rate of $48$ kHz, and block sizes were varied from $16$ to $4096$ samples in powers-of-two increments.

Audio routing and real-time scheduling were managed using the JACK Audio Connection Kit \cite{jack2025}, a Linux sound server optimized for low-latency audio processing. Latency measurements were obtained using the \texttt{jack\_delay} utility, which estimates round-trip latency by analyzing frame timing between the input and output ports of the JACK server.

\begin{table}[ht]
\centering
\def~{\hphantom{0}}
\caption{Buffer latency and corresponding measured end-to-end latency for various block sizes at a sampling rate of $48$ kHz.\label{tab:end2end_latency}}
\begin{tabular}{@{}ccc@{}}
\toprule
Block size & Buffer latency (ms) & \makecell{End-to-end latency (ms)}\\
\midrule
~~16  & ~0.33  & --     \\
~~32  & ~0.67  & --     \\
~~64  & ~1.33  & ~~7.13 \\
~128  & ~2.67  & ~12.26 \\
~256  & ~5.33  & ~19.47 \\
~512  & 10.67  & ~26.13 \\
1024  & 21.33  & ~47.47 \\
2048  & 42.67  & ~90.13 \\
4096  & 85.33  & 175.70 \\
\bottomrule
\end{tabular}
\end{table}

\Cref{tab:end2end_latency} presents the measured end-to-end latency values alongside the corresponding buffer latency for each block size. Measurements were performed for both single-channel and multichannel configurations up to $24$ channels. No significant variation in end-to-end latency was observed for different channel counts. For block sizes below 64 samples, stable audio processing could not be maintained due to latency constraints in the operating system and audio interface. For block sizes of $64$ samples and above, end-to-end latency increased with buffer size, ranging from $7.13$ ms at $64$ samples to $175.70$ ms at $4096$ samples.

These results highlight the need to minimize processing latency, as it limits the smallest stable block size. Since end-to-end latency scales with buffer size, efficient processing is required to maintain low-latency performance without compromising stability, especially in interactive applications where responsiveness is desired.

\section{Conclusion}
\label{sec:gpu2025:conclusion}

\noindent This paper presented a real-time implementation of a \gls{gpu}-accelerated loudspeaker-based auralization system for the interactive synthesis of large room acoustics with integrated feedback cancellation. The system was designed to support scalable loudspeaker-based synthesis with practical flexibility in filter length and channel count. The implementation was evaluated in terms of processing time by comparing the performance of the \gls{gpu} and \gls{cpu} implementations. It was shown that the \gls{gpu} implementation achieves a performance gain of up to $60$ times compared to the \gls{cpu} implementation, enabling real-time auralization at $48$ kHz with a block size of $128$ samples for synthesis filters with a duration of up to $20$ seconds and $32$ output channels. This block size corresponds to a buffer latency of $2.67$ ms, and an end-to-end latency of $12.26$ ms as measured using an \textit{RME Fireface UFX+} audio interface connected to the auralization system. The results demonstrated that the implementation supports real-time auralization with minimal latency, making it suitable for interactive applications with large room acoustics. While the Python implementation of the system was not optimized for real-time performance, it was shown to achieve real-time performance using accelerated libraries. Future work will focus on further optimizing the system for real-time performance, and on interactive perceptual evaluation of the system.

\section{Acknowledgments}
This research work was carried out at the ESAT Laboratory of KU Leuven, in the frame of KU Leuven internal funds C3/23/056 “HELIXON: Hybrid, efficient, and liquid interpolation of sound in extended reality” and C14/21/075 "A holistic approach to the design of integrated and distributed digital signal processing algorithms for audio and speech communication devices”, FWO SBO Project The sound of music - Innovative research and valorization of plainchant through digital technology" (S005319N), FWO Large-scale research infrastructure "The Library of Voices - Unlocking the Alamire Foundation's Music Heritage Resources Collection through Visual and Sound Technology" (I013218N), and SBO Project "New Perspectives on Medieval and Renaissance Courtly Song" (S005525N). The research leading to these results has received funding from the Flemish Government under the AI Research Program and from the European Research Council under the European Union's Horizon 2020 research and innovation program / ERC Consolidator Grant: SONORA (no. 773268). This paper reflects only the authors' views and the Union is not liable for any use that may be made of the contained information.

\bibliographystyle{ieeetr}

\end{document}